\documentclass{emulateapj}

\usepackage{natbib,amsmath}
\usepackage{graphicx}
\usepackage{hyperref}
\usepackage{float}
\usepackage{subfigure}

\newcommand{\forloop}[5][1]%
{%
\setcounter{#2}{#3}%
\ifthenelse{#4}%
	{%
	#5%
	\addtocounter{#2}{#1}%
	\forloop[#1]{#2}{\value{#2}}{#4}{#5}%
	}%
	{%
	}%
}%

\newcommand{\ctbd}[1]{}

\newcommand{\ccdsize}[1]{\ensuremath{\rm #1\times\rm#1}}
\newcommand{\fovsize}[2]{\ensuremath{\rm #1 #2\times\rm#1 #2}}

\newcommand{\pxs}{\ensuremath{\rm \arcsec pixel^{-1}}}

\newcommand{\reffig}[1]{Fig.~\ref{fig:#1}}

\begin{document}
\title{Precision multi-band photometry with a DSLR camera}

\author{
	M.~Zhang\altaffilmark{1},
	G.~\'A.~Bakos\altaffilmark{1,2,3},
    K.~Penev\altaffilmark{1},
    Z.~Csubry\altaffilmark{1},
    J.~D.~Hartman\altaffilmark{1},
    W.~Bhatti\altaffilmark{1},
    M.~de Val-Borro\altaffilmark{1}
}
 
\altaffiltext{1}{Department of Astrophysical Sciences, 
	Princeton University, Princeton, NJ 08544; 
	email: mzzhang2014@gmail.com}
\altaffiltext{2}{Sloan Fellow; email: gbakos@astro.princeton.edu}
\altaffiltext{3}{Packard Fellow}


\begin{abstract}

Ground-based exoplanet surveys such as SuperWASP, HATNet and KELT have
discovered close to two hundred transiting extrasolar planets in the
past several years.  The strategy of these surveys is to look at a
large field of view and measure the brightnesses of its bright stars to
around half a percent per point precision, which is adequate for
detecting hot Jupiters.  Typically, these surveys use CCD detectors to
achieve high precision photometry.  These CCDs, however, are expensive
relative to other consumer-grade optical imaging devices, such as
digital single-lens reflex cameras (DSLRs).

We look at the possibility of using a digital single-lens reflex camera
for precision photometry.  Specifically, we used a Canon EOS 60D camera
that records light in 3 colors simultaneously.  The DSLR was integrated
into the HATNet survey and collected observations for a month, after
which photometry was extracted for 6600 stars in a selected stellar
field.

We found that the DSLR achieves a best-case median absolute deviation
(MAD) of 4.6\,mmag per 180\,s exposure when the DSLR color channels are
combined, and 1000 stars are measured to better than 10\,mmag (1\%). 
Also, we achieve 10\,mmag or better photometry in the individual
colors.  This is good enough to detect transiting hot Jupiters.  We
performed a candidate search on all stars and found four candidates,
one of which is KELT-3b, the only known transiting hot Jupiter in our
selected field.  We conclude that the Canon 60D is a cheap, lightweight
device capable of useful photometry in multiple colors.

\end{abstract}

\section{Introduction}

The last two decades have been very exciting for exoplanetary science. 
Since 1995, when 51 Pegasi b became the first planet discovered around
a main-sequence star, over 1000 exoplanets have been discovered, and
thousands of candidates are waiting to be confirmed.

So far, the two most common ways to detect exoplanets are the transit
and radial velocity methods, respectively.  The former looks for dips
in a starlight due to planets passing in front of them.  The latter
uses spectroscopy to look for changes in the radial velocity of the
star as it orbits around the barycenter of the planetary system.  The
most well-known project using the transit method is the \textit{Kepler
Space Mission}, which examined 100 square degrees of the sky, and
measured the brightness of 150,000 stars to a precision high enough to
detect Earth-size planets near the habitable-zones of Sun-like stars in
the best cases.  Most of the stars observed by \textit{Kepler} are
necessarily very faint, and therefore not easy targets for follow-up
observations.

Ground-based transit searches have been operating before, during, and
after \textit{ Kepler}, using the same principle, but not achieving the
ultra-high-precision photometry possible from space.  Their strategy is
to take wide-field images of the night sky and measure the bright stars
within using commercially available CCDs.  With the exception of a few
hot Neptunes, such as HAT-P-11b \citep{hat_p_11b}, HAT-P-26b
\citep{hat_p_26b} and HATS-8b \citep{hats_6b}, this strategy is only
precise enough to detect hot Jupiters.  However, because the surveys
monitor a large number of fields at multiple sites over many years,
they routinely detect hot Jupiters around bright stars that are
suitable for follow-up observations.  Exames of ground-based transit
searches include HATNet \citep{hatnet_2002,hatnet}, SuperWASP
\citep{superwasp}, KELT \citep{kelt}, and HATSouth \citep{hatsouth},
which have collectively discovered and characterized close to two
hundred transiting exoplanets.

HATNet, in particular, was established in 2003, and consists of 6 fully
automated observatories at two sites: Arizona and Hawaii.  Each
observatory has a 11\,cm f/1.8 lens in front of a \ccdsize{4K}
front-illuminated CCD, giving a \fovsize{10.6}{\arcdeg} field of view. 
Every night, the mini-observatories use a weather sensor to measure
wind speed, humidity, cloud cover, precipitation, and nearby lightning
strikes.  If the weather is appropriate, the domes open and the
telescopes use their CCDs to take images of designated fields for the
entire night.  HATNet now monitors more than 100,000 stars at better
than 1\% precision every year, for a lifetime total of close to 1
million such stars, and has discovered more than 50 transiting
exoplanets.

The CCDs that HATNet and other surveys use typically cost at least
10,000 USD, which is a significant obstacle against the creation of new
exoplanet surveys, or the expansion of existing ones.  For this reason,
we investigate the possibility of using a mid-range, commercial digital
single-lens reflex (DSLR) camera for precision photometry.  These
devices have excellent electronic characteristics for their price,
which is around 1000 USD, and are more affordable for the public. 
Furthermore, they have improved drastically over the past several
years, and consumer demand continues to force manufacturers to produce
better products.

Several authors have already tested the use of DSLRs for photometry. 
\cite{hoot_2007} discusses the basic characteristics of DSLRs, then
used a Canon EOS 350D to perform absolute photometry by transforming
instrumental magnitudes to magnitudes in standard Landolt filters. 
They found RMS errors of 0.347, 0.134, and 0.236\,mag for the B, V, and
R bands respectively.  \cite{littlefield_2010} used the 12-bit Canon
20Da camera, and measured the transit of exoplanet HD 189733b, whose
star is at V=7.67, to around 6 mmag accuracy.  The author achieved this
accuracy by using a high ISO setting and extreme defocusing, so that
the star image was around 35 pixels in diameter.  This defocusing
improves photometry through reducing intra- and inter-pixel variations,
but is not suitable for exoplanet surveys because of stellar crowding,
and because when faint stars are defocused, photon noise from the sky
background, read noise, and dark noise start to dominate.

\cite{kloppenborg_2012} took 500 DSLR images in groups of ten,
12-second exposures and measured the brightness of stars with $3.5 < V
< 7.5$ to a mean uncertainty of 0.01 mag, and stars with $7.5 < V <
8.0$ to a mean uncertainty of 0.02 mag or better.  They used a Canon
450D and defocused so that stellar images were 10-15 pixels in
diameter, and used the Tycho catalogue to account for stellar blending
in aperture photometry.  Unfortunately, there are two aspects of this
process that are not ideal for the purposes of a wide-field photometric
survey.  One of them is defocusing, mentioned earlier; the other is the
12-second exposure time.  Our 60D has a mean shutter lifetime of
100,000 exposures (according to the Canon website), and at 12 seconds a
frame, it would last on the order of two months before failing.

Finally, \cite{guyon_2012} used a Canon 550D and 500D to observe the
star HD 54743 without defocusing, by recombining the 4 color channels
into a single image, and choosing comparison stars with a similar PSF
as that of the target.  This step implicitly selects a comparison star
with similar color, optical aberrations, and position on the Bayer
filter.  With regular aperture photometry, they achieved a precision of
around 10\%.  With the PSF selection technique, they achieved 2\% in
the green channel, 2.5\% in the red, and 3.5\% in the blue, which is
close to the photon noise limit.  However, it must be noted that they
observed a relatively faint star during bright time, and in this work
they did not test the accuracy of their system during better observing
conditions.

\subsection{DSLR vs CCD}

There are a number of advantages and disadvantages of using DSLRs
instead of CCDs.  The disadvantages are severe.  DSLRs have lower
quantum efficiency and a smaller full well capacity---both of which
increase the Poisson noise of detected starlight, thus imposing a
strict upper limit on the photometric precision.  DSLRs are generally
not cooled, and dark current increases dramatically with temperature. 
This increases the dark noise at high ambient temperatures and makes
dark frames a less accurate reflection of the dark-current signature of
image frames.  Additionally, DSLRs have a Bayer filter superimposed on
top of their sensors.  This makes aperture photometry difficult because
as the star moves across the image due to tracking or pointing errors,
a varying fraction of its light falls onto a pixel of a given color. 
Thus, the brightness of a constant star appears to vary by as much as
13\%, as indicated by \cite{guyon_2012}.  Finally, as we will show in
this paper, cameras like the Canon EOS 60D perform post-processing on
images even when writing in ``RAW'' format.  There is no documentation
of this post-processing, and frames exhibit a variety of disturbing
behavior that we have no explanation for.

The primary \textit{ advantage} of a DSLR is its low cost.  The Canon EOS
60D that we use is around 700 USD as of writing.  By comparison, a CCD
camera with a \ccdsize{4K} Kodak KAF16803 chip is $\sim$10,000 USD, and
even a lower-end CCD is a few thousand dollars.  Additionally, DSLRs
require no external device for focusing; a DSLR setup is cheaper and
more lightweight, making it more accessible for amateur astronomers. 
Finally, the Bayer filters of DSLRs allow for fully simultaneous
photometry at three colors, which is very difficult for CCDs, though
can be done through the use of expensive dichroic beam splitter systems
\citep{grond}.

\section{Hardware setup}

\subsection{Initial tests}

We attached the Canon 60D camera to a Fornax F50 telescope mount on the
roof of Peyton hall at Princeton University's campus.  We observed
selected stellar fields by simple tracking of the mount, without
autoguiding.  Focusing was performed using a very rudimentary
algorithm.  Most features of the camera, such as USB connectivity,
external power supply, communication from Linux, were all tested from
Peyton hall's roof.  While the astro-climate from Princeton is
sub-optimal (heavy light pollution, small fraction of clear skies),
these tests on real stars over multiple nights helped prototype
operations.  Once we reached stable operations, we relocated the camera
to the Fred Lawrence Whipple Observatory in Arizona.

\subsection{HAT10}
\label{subsec:hat10_hardware}

HAT10 is a telescope unit from the HAT Network of Telescopes (HATNet),
located at the Fred Lawrence Whipple Observatory (FLWO) of the
Smithsonian Astrophysical Observatory (SAO).  A HAT unit is a fully
automated observatory using a CCD and a telephoto lens on a common
mount, all within a small clamshell dome, which is equipped with a rain
sensor and a photosensor that force the dome to close during rain or
daytime.  The unit is also connected to a weather sensor in a control
building which measures clouds, lightning, and wind speed.  All devices
are controlled via a single Linux PC\@.  Every night, if the sensors
detect favorable conditions, the dome automatically opens, the mount
tracks a certain field, and the camera continuously takes 3-minute
frames of the field.  At dawn, or when the weather deteriorates, the
dome automatically closes, and the HAT goes to sleep.

\begin{figure}[h]
\includegraphics[width= 0.5\textwidth]{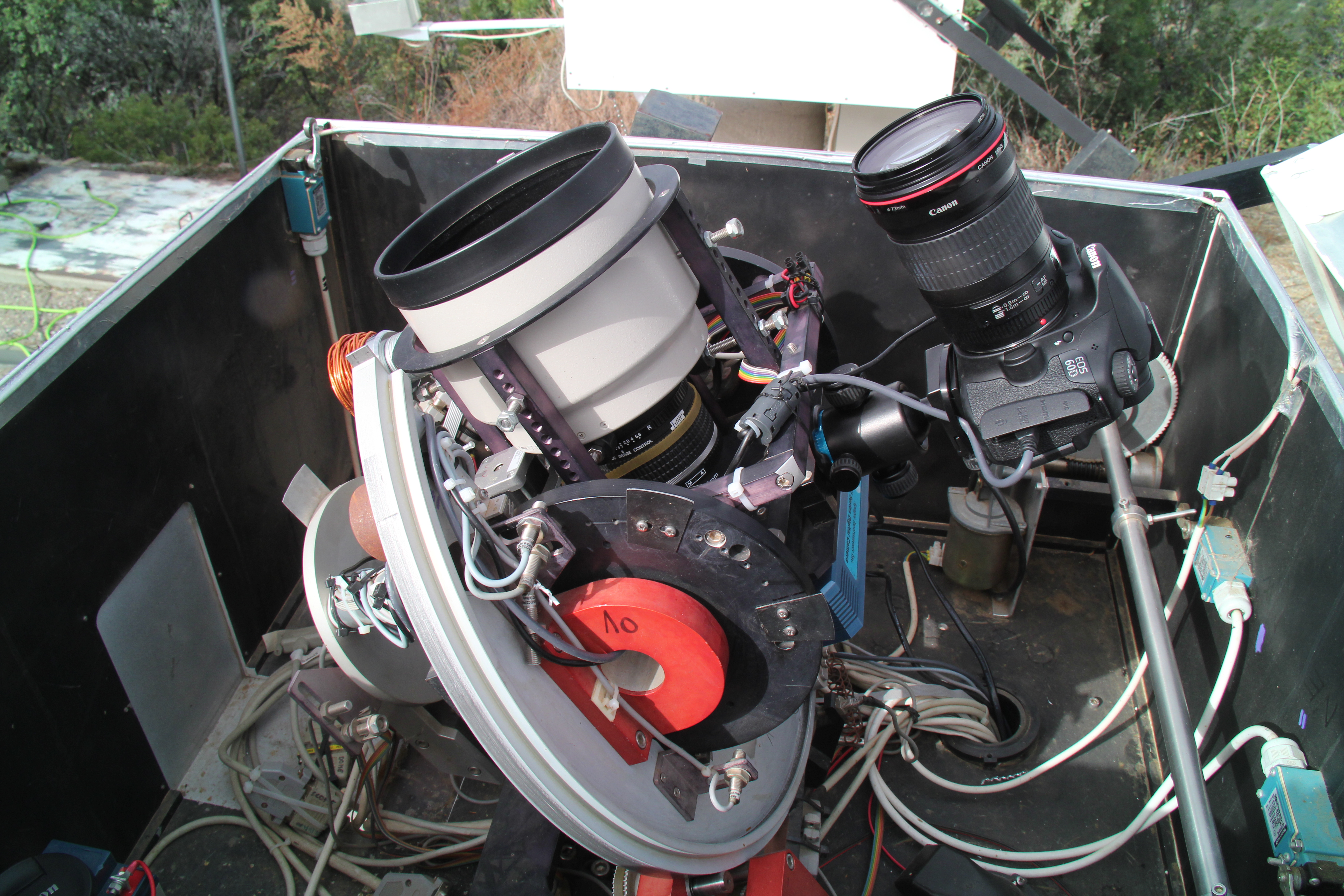}
\caption{
	The interior of the HAT10 telescope unit with the horseshoe mount
	holding the blue square-shaped CCD, a telephoto lens, and large
	dewcap.  Piggybacked on the right side is the DSLR camera with a
	Canon 135mm f/2 lens.
}
\label{fig:hat10_interior}
\end{figure}

The DSLR was attached to the HAT10 mount and was pointing in the same
direction as the CCD\@.  The mechanical installation was done in 2014
February with a ballhead and custom-made converters
(\reffig{hat10_interior}).  The DSLR camera is a Canon EOS 60D, a
midrange DSLR often used for astrophotography.  We used a Canon EF
135mm f/2L lens at the f/2 setting.

\subsection{Canon 60D Camera Properties}

A number of experiments were performed to characterize the DSLR camera. 
Additionally, we have tried to look for official documentation on these
properties; unfortunately, none seems to be public.  We have also
searched for other measurements of camera properties, but the only
measurements that exist were found on personal websites such as
clarkvision\footnote{
\url{http://www.clarkvision.com/photoinfo/evaluation-canon-7d/index.html}}.
The information below is likely the most detailed publicly available
description of the properties of the 60D.  

\subsubsection{General Properties}
\begin{figure}[h]
\centering
\includegraphics[width=0.2\textwidth]{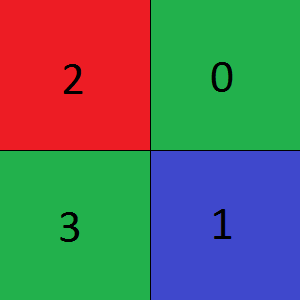}
\caption{
	Bayer filter of the 60D, as viewed by an observer on top of the
	lens looking downwards at the filter.  The horizontal dimension in
	this figure represents the horizontal dimension of the sensor, and
	similarly for the vertical dimension.  Each square represents a
	$4.3\mu$ pixel.  The numbers represent the ``color channels'',
	which is a convention we use in our data analysis, and in the text.
}
\label{fig:bayer_filter}
\end{figure}

At full resolution, the Canon EOS 60D takes $5184\times3456$ pixel
images with its CMOS sensor, for a total of 18 megapixels.  Each pixel
has a physical size of 4.3 by 4.3 micrometers.  With our Canon 135 f/2
lens, the camera has a field of view of $9.6\arcdeg\times6.4\arcdeg$,
corresponding to a pixel scale of 13.3\,\pxs.
The camera has a Bayer filter on top of the CMOS sensor, laid out in
the pattern indicated in \reffig{bayer_filter}.  Each square in the
Figure represents one pixel.

To break the ambiguity between the two green channels, we will refer to
the channels by the numbers 0 (green), 1 (blue), 2 (red) and 3 (green),
whose correspondence to the Bayer filter is shown in
\reffig{bayer_filter}.  These numbers are mostly arbitrary, and were
originally chosen because they were the order in which {\tt dcraw},
which we used to convert RAW files to FITS, writes the channels.

When shooting in RAW mode (as compared to producing compressed JPEG
files), the camera outputs 4-channel data at 14 bits per pixel, each
channel corresponding to a color.  Pixel values have an offset of 2048,
a value which represents zero light.  This offset is non-zero to
prevent clipping negative values to zero or dedicating an entire bit to
indicate the sign of the value.

\begin{figure}[h]
\includegraphics [width= 0.5\textwidth]{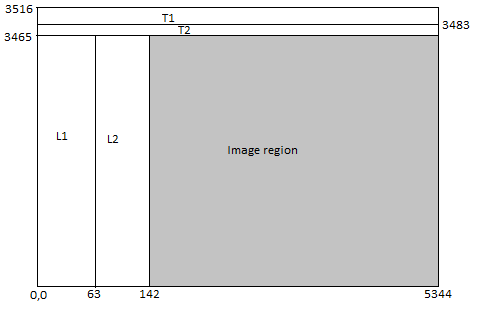}
\caption{
	Layout of the entire CMOS sensor of the Canon 60D camera.  The
	regions are not to scale.  Aside from the image region, there are
	at least 4 light-insensitive regions: a 63-pixel strip at the very
	left, a 79-pixel strip next to it, a 33-pixel strip at the very
	top, and a 18-pixel strip immediately below it.  Origin of the
	coordinate system is at the lower-left corner of the lowest and
	left-most pixel; $x$ increases to the right and $y$ towards the
	top.  Note that this convention is different from that of GIMP,
	Adobe Photoshop, and most other image editors, and has a 0.5-pixel
	offset from the convention used by the ds9 FITS viewer software.
}
\label{fig:sensor_layout}
\end{figure}

Not all of the chip pixels are sensitive to light.  Based on our own
measurements with this camera we produced the diagram shown in
\reffig{sensor_layout}, describing the layout of the different regions
on the chip.  We have failed to find any public documentation
describing this layout.

Significant effort was spent trying to determine the behavior of these
dark regions in the hopes of exploiting them to remove offsets or
fixed-pattern noise.  We discovered that all regions except T1 are
almost indistinguishable in bias frames.  For dark frames, L2 and T2
are almost indistinguishable from the image region, while L1 and T1 are
clearly distinct and less bright.  For this reason we suspected that L1
is a bias region, where the pixels are exposed for 0 seconds, while L2
and T2 are dark regions, where the pixels are covered with a strip of
metal and exposed for the same length as the image region.  However,
further research showed that this hypothesis is incorrect.  Moreover,
the non-image regions exhibit weird and unexplained behavior, such as
channel-dependent mean intensity even though the region receives no
light.  The behavior of the image region is consistent and generally in
accordance with expectations.  Therefore, we simply crop out the image
region in all of our processing and do not use the other regions.  This
is an area of further possible improvement.

\subsection{Electronic properties}
\begin{table}
\begin{center}
$
\begin{array}{lr}
\hline
\text{Property} & \text{Value} \\ 
\hline
\text{Gain (e-/ADU)} & 2.10\\
\text{Saturation (ADU)} & 13,585\\
\text{Saturation (e-)} & 24,200\\
\text{Read noise (e-)} & 15.5\\
\hline
\end{array}
$
\end{center}
\caption{
Electronic properties of the 60D at ISO100 sensitivity setting, as
measured in the lab.
}
\label{table:electronic_properties}
\end{table}

Table \ref{table:electronic_properties} shows the electronic properties
of the 60D at ISO100, the setting we used.  We measured all of these
properties in our lab.  Read noise, for example, was measured by taking
pairs of very short exposures in a dark room.  Gain was measured by
taking pairs of exposures of a uniform white light at different
exposure times and calculating the noise as a function of the mean
intensity.  Here we assumed that the noise consists of photon noise
plus read noise.  Saturation was measured by taking an image of a
bright source.

\subsubsection{Linearity}
\begin{figure}[h]
\includegraphics [width= 0.5\textwidth]{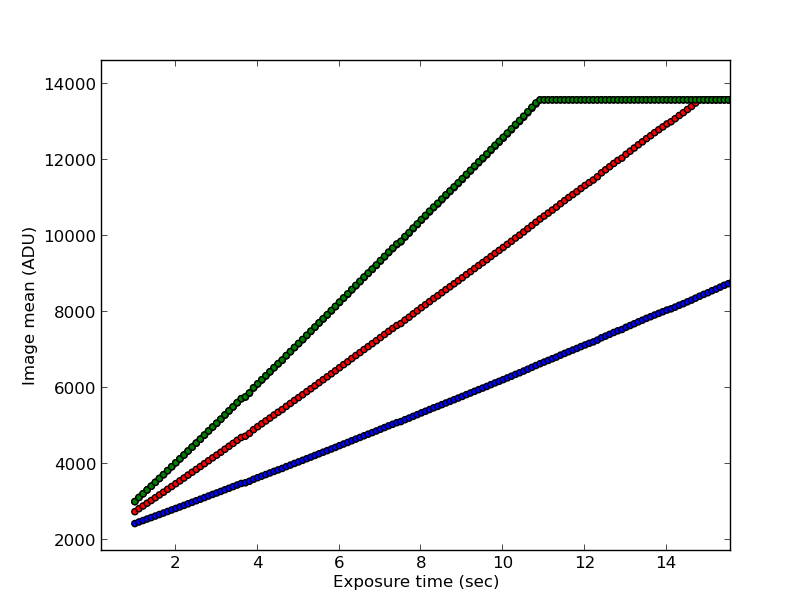}
\caption{
	Mean of all image pixels versus exposure time for all 4 channels at
	ISO100.  Both green channels are plotted, but one is hidden
	underneath the other.
}
\label{fig:linearity}
\end{figure}

The Canon EOS 60D exhibits almost perfect linearity all the way to the
saturation value in all channels, as shown in \reffig{linearity}.  This
plot was produced by putting 3 pieces of paper on top of the lens of
the lab camera, with the lab lights turned on.  This setup yielded
$\sim$1000 ADU/s intensity.  We then took exposures of increasing
duration, ranging from 1 to 15 seconds.  The two small kinks in the
lines in \reffig{linearity}, at 4\,s and 7\,s, could be due to actual
changes in lighting rather than camera properties.

To characterize the linearity, linear regression was used to fit the
mean ADU as a function of exposure time.  Points with means at the
saturation value were rejected.  The resulting $1- r^2$ (the percentage
of variance not explained by the best-fit line) is
$1.1\times10^{-4}$ for channel 0, 
$3.3\times10^{-4}$ for channel 1, 
$1.9\times10^{-4}$ for channel 2, and 
$1.1\times10^{-4}$ for channel 3.

The linearity is less perfect at a few hundred ADUs from the saturation
level, but as shown by the r values above, it is still nearly perfect
for nearly the entire dynamic range.

\subsubsection{Dark current}

Dark current is caused by thermal electrons that jump the bandgap from
the valence to the conduction band of the semiconductor.  The general
formula describing dark current is $D_{ADU} = \alpha \cdot
exp(E_g/(kT))$, where $E_g=1.1 eV$ is the bandgap of silicon (the main
material used in CMOS and CCD photodiodes) and $\alpha$ is a
chip-dependent constant.

For a CCD, the dark current causes an increase in pixel values as
exposure time is increased.  This does not happen for this particular
DSLR (and probably is a general feature of DSLR cameras).  We took test
frames at ISO800 setting using exposures times of 1, 2, 4, 8, and 16
minutes, and saw only a 3 ADU increase in mean pixel values from 1 to
16 minutes.  The increase is not linear.  This is because the 60D have
a built-in mechanism to suppress dark current, even when the
``dark-noise suppression'' option of the camera is turned off.  We have
not been able to find any documentation about how this suppression
works.

To determine the dark current for the HAT10 camera, we look at the dark
noise instead.  We expect a Poisson noise characteristic.  We plotted
the image variance of every dark frame against the Boltzmann factor
$e^{-\frac{E_g}{kT}}$, where T is from the sensor temperature, as read
out by our control software.  The expected variance is given by:
\begin{align*}
V_{ADU} = \frac{D_{ADU}}{g} + R_{ADU} = \frac{D_{e-}}{g^2} + R_{ADU}
\end{align*}
where g is the gain, $D_{ADU}$ is the dark current in ADU per 180
seconds, R is the readout variance, and V is the total variance.

\begin{figure}[h]
\includegraphics[width= 0.5\textwidth]{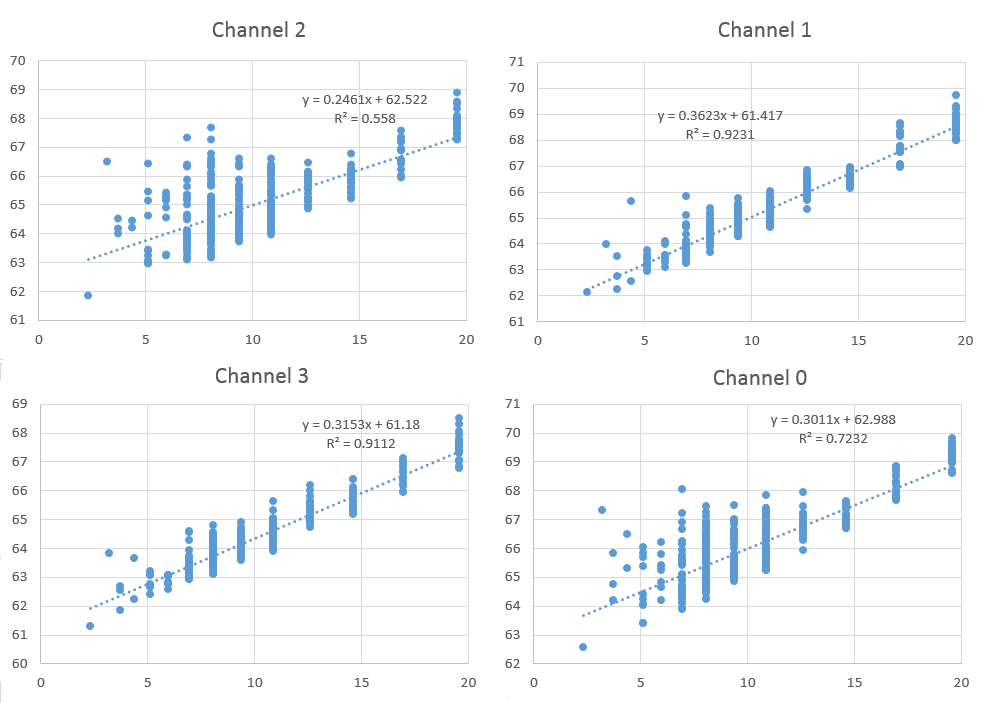}
\caption{
Variance in $ADU^2$ versus $10^{20}\text{exp}(-E_g/kT)$ for dark frames
taken by HAT10 for all four channels, along with best-fit lines and
their parameters.
}
\label{fig:variance_vs_temp}
\end{figure}

\reffig{variance_vs_temp} shows the variance as a function of the
Boltzmann factor.  The best fit is:
\begin{align*}
D_{e-} &\approx 1.4 \times 10^{20} e^{-E_g/kT}\\
R_{ADU} &\approx\, 62\, ADU^2
\end{align*}
corresponding to around 0.1 electrons per second at $20^{\circ}C$. 
Note that the dark current depends on channel: channel 0 and 2 have
much larger dispersion with respect to the best-fit line than the other
channels, even though these are dark frames.  This is not consistent
with light leakage, because the two green channels (0 and 3) are
significantly different.  This figure also shows that the dark noise
due to dark current is low compared to readout noise for temperatures
around $20^{\circ}C$ and exposure times of a few minutes.

\section{Observations}
The hardware setup was described in more detail in Section
\ref{subsec:hat10_hardware}.  We carried out simultaneous observations
with the Canon 60D plus Canon 135mm f/2 lens, and the Apogee U16m CCD
plus its Canon 200mm f/1.8 FD lens with sloan $r$ filter (the default
HATNet setup).  The Canon 60D system was piggy-backed next to the
Apogee CCD, and was taking simultaneous exposures of the same length. 
This allows a good comparison to be done between the CCD and the DSLR.

After a few days of debugging, the DSLR, under the control of a Linux
computer, started taking useful data on February 22 and continued doing
so until May 30.  Shortly after sunset, if the weather was favorable,
the dome opened and both the CCD and the DSLR began taking flat-fields. 
After the sky became dark, the mount pointed to a selected stellar
field on the sky and the DSLR started taking three-minute exposures. 
Whenever the mount first points to a new field, it takes `pointing'
frames which are used to determine the true field center and correct
for initial pointing errors.  Following the pointing frames, the camera
took 180-second science object frames.  After every ninth exposure, it
took a 30-second focus frame, and any deviations from the target focus
were corrected.  The focusing procedure is described in Section
\ref{subsec:focusing}.

Once the field was near setting (below an altitude of 33 degrees,
corresponding to an airmass of 1.8) or the Moon was too close to the
field, the next highest-priority stellar field was chosen.  Around
sunrise, the camera took flat fields again before the dome closed and
the HAT unit went to sleep for the day.

\subsection{Camera settings}
Certain camera settings are essential for scientific photometry.  We
shoot in RAW format to avoid additional noise due to the lossy
compression to JPEG format.  We set the image size to be the largest
possible.  We turn the long-exposure noise reduction feature of the
camera off, since in this mode the camera would take a dark frame of
the same length as our observation after each frame and would thus
spend half the observing time without collecting photons.  The camera
is turned to Bulb mode and the lens is set to Autofocus, so that the
computer can control both the exposure time and focus, respectively.

There are four camera parameters that can be changed to optimize
photometry for a certain magnitude range: ISO (sensitivity), aperture,
exposure time, and focus.

Increasing ISO decreases the gain.  Since saturation ADU does not
increase at ISOs higher than 100 (which is the lowest possible
setting), this also decreases the number of electrons that can be
represented before reaching the saturation limit, thus decreasing the
magnitude range of stars whose brightness can be precisely measured. 
(Due to
anti-blooming features implemented on the chip, charges are simply lost
after saturation; they do not overflow to neighboring pixels, and thus
can not be measured to provide precise photometry.) However,
the electron-equivalent read noise is $Var_{e} = (gR_1)^2 + R_0^2$,
where g is the gain, $R_1$ is the pre-amplifier noise component, and
$R_0$ is the post-amplifier noise component.  The read noise therefore
\textit{ decreases} with the ISO setting and approaches $R_0 = 2.62 e^-$ at
high ISO\@.  Therefore, to optimize for any given star, it is always
better to use the highest ISO possible without reaching saturation for
a given exposure time.  To survey the brightest stars optimally, it is
best to use ISO100\@.

The considerations above indicate that ISO100 or 200 is probably the
best ISO\@.  The former is better for higher dynamic range, the latter is
better for lower readout noise.  While we were not fully aware of all
of these considerations at the time of setup, the setting we have been
using (ISO100) turned out to be the optimal choice.

\subsection{Stellar Fields}

The observed field during this test is named G180 in the HATNet
nomenclature.  It is centered at RA=144.6 degrees and Dec=37.4 deg. 
This field was selected because it is high above the horizon after
sunset in February, and because it contains the transiting exoplanet
KELT-3b.  The field is in between the Lynx and Leo Minor
constellations.  G180 was also observed by the other HAT telescopes,
and thus it is a good control field for comparing the results.

When field G180 was setting, the camera observed another field, called
G113, until the dawn.  This field was extremely dense in stars due to
its proximity to the plane of the Milky Way, and is centered very close
to the `head' of Draco.  Data for G113 have not been fully analyzed,
and are not subject to this paper.

\subsection{Calibration Frames}

Astronomical CCDs can easily take dark frames, by simply exposing
without opening the shutter.  Usually, HATNet CCDs take dark frames at
the end of every night.  Unfortunately, the Canon 60D has no built-in
ability to take frames without opening the shutter.  We investigated
whether firmware modifications such as Magic Lantern
(\url{http://magiclantern.fm}) are capable of introducing this ability
but found nothing relevant.  For this reason, we only took 20 bias
frames (on the night of March 6), by covering the lens, and taking very
short exposures.

Darks were taken in two ways.  First, whenever the weather was bad
during night time, the dome remained closed, and the camera took
180-second exposures of the inside of the dome.  Second, on the night
of March 5, the lenscap was manually put on the lens, and the camera
took dark frames the entire night.  A total of 1040 darks were taken
during bad weather, while 125 were taken on March 5 with the lens cap
on.

Skyflats were taken during both evening and morning twilight, depending
on weather.  In the morning, flat-fields start with 180 s of exposure
time and remain at that exposure until the average pixel value exceeds
the target value of 2/3rd of the saturation value.  After that, the
exposure time is gradually reduced (based on the intensity values of
all previous frames) in order to keep the image brightness constant, at
2/3rd of the saturation value.  We stop taking flat fields when the sky
becomes too bright, which corresponds to an exposure time of 1 second. 
In between flat fields, the mount points to slightly different areas of
sky.  This allows for stars to be removed from the master flat, which
is derived by median averaging the individual skyflat frames.

\subsection{Focusing}
\label{subsec:focusing}

To obtain high precision photometry with a DSLR, most authors choose to
defocus the lens, with the exception of \cite{guyon_2012}.  Defocusing
increases the PSF width, thereby mitigating the effect of the subpixel
structure and the Bayer filter on the results.  However, defocusing has
many problems that make it unsuitable for an automated photometric
wide-field survey.  Increasing the PSF makes stellar blending worse and
increases the photon noise due to the sky background.  It makes the PSF
non-Gaussian, highly variable, and elongated at the edges, thus
violating the assumptions made by most source finder and flux extractor
software, including our own reduction pipeline of HATNet.  If the PSF
becomes donut-like, star detectors may spuriously detect multiple peaks
and count them as multiple stars, which confuses autofocusing and
astrometry algorithms.  Instead of defocusing, we choose the opposite
approach, i.e.~to focus as perfectly as possible, and continuously
monitor stellar PSFs to keep focus in the same position.

The Canon EF 135 mm lens has built-in focusing ability that can be
controlled by a computer.  We note that the built-in \textit{ autofocusing}
capability of the Canon 60D camera does not work under low light
conditions, and on stars.  On February 18, shortly after the DSLR was
installed, a focus series was taken during the night.  Starting at
focus position 45 (in arbitrary units), an image was taken at every
other focus position until we reached 70.  The focus position refers to
the number of quanta of focus adjustments from the farthest possible
focus, which is beyond infinity.  The image taken at each focus
position was split into the 4 channels, corresponding to the 4 colors. 
Each channel was, in turn, split into 9 regions: upper left, upper mid,
upper right, middle left, center, etc.  A star detector was run on
every channel, and every star was fit with a four-parameter rotated
elliptical Gaussian of the form:
\begin{align*}
f(x,y) = A\text{exp}(-\frac{1}{2}[S(x^2 + y^2) + D(x^2 - y^2) + K(2xy)])
\end{align*}
Here, the star is assumed to be centered on (0,0); in reality, the
center of the star is estimated by computing the centroid of the light. 
$S$ is related to the inverse of the square of standard deviation, and
represents ``sharpness''.  $D$ represents ellipticity in the $x$,$y$
direction, while $K$ represents the ellipticity in the diagonal
directions.

\begin{figure*}[t]
\begin{center}
\subfigure[
	S (top), D (middle), K (bottom) parameters of the stellar PSF
	vs.~focus position for channel 0 (green bandpass).  The nine curves
	in each panel represent the nine different image regions (they are
	not related to the RGB colors of the DSLR).
	\label{fig:focseries_ch0}]
	{\includegraphics[width=0.45\textwidth]{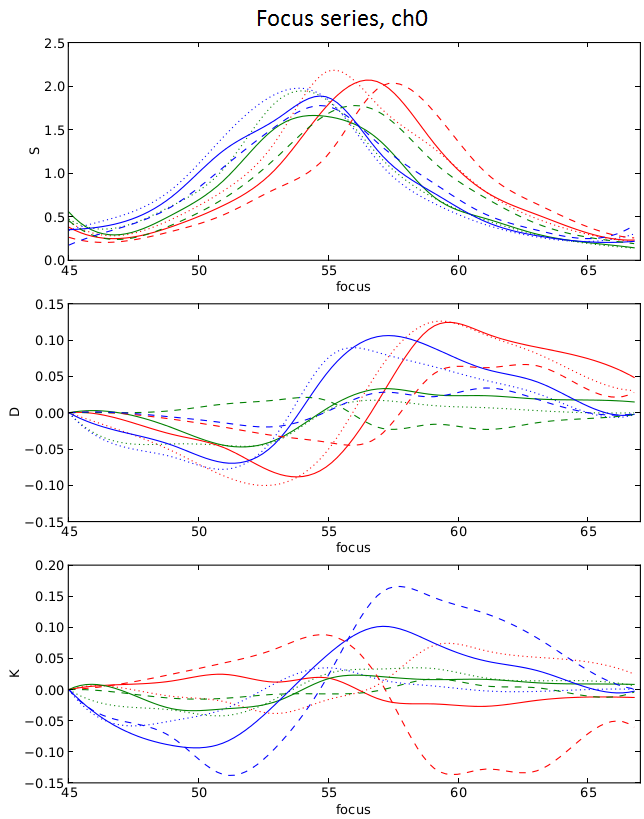}}\qquad
\subfigure[
	S,D,K vs.~focus position for channel 1 (red bandpass). Note that
	the peak value of S is smaller (meaning a wider Gaussian profile),
	and the position of the maximum is at a different location than in
	the green bandpass.  \label{fig:focseries_ch2}]
	{\includegraphics[width=0.45\textwidth]{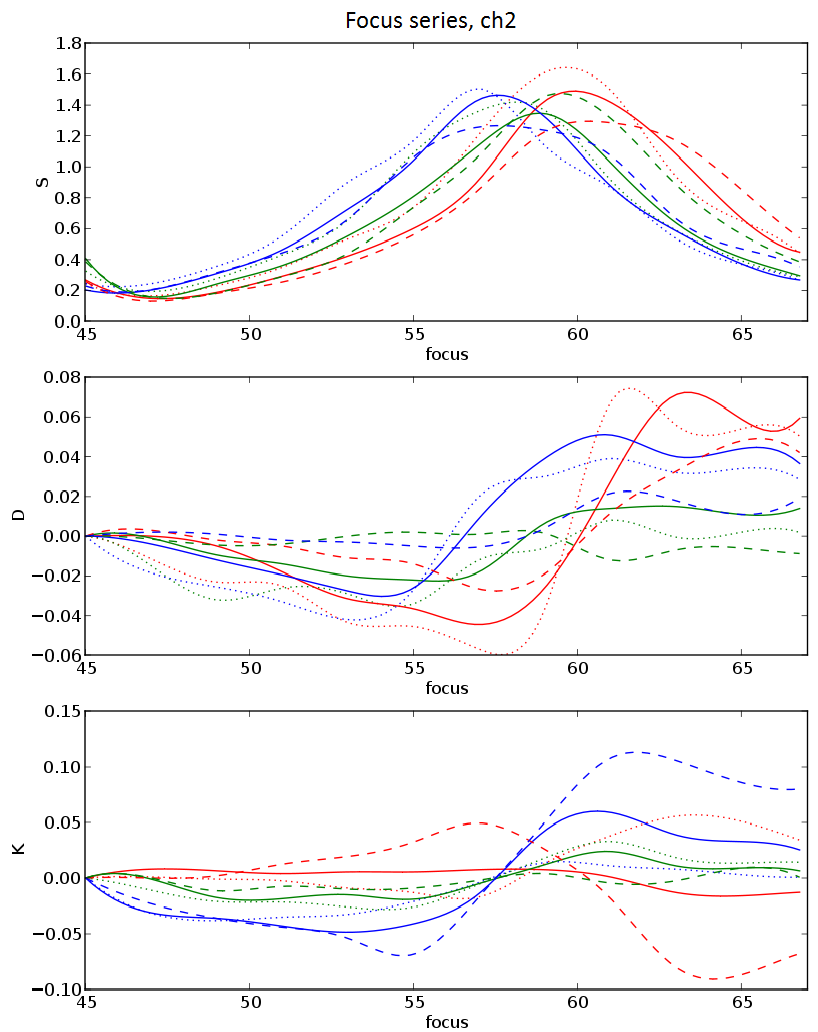}}
\caption{
	Focus curves for the green (left) and red (right) channels.  Notice
	the differences in optimal focus between the two channels.  Not
	surprisingly, channel 3 (not shown) is very similar to 0; somewhat
	surprisingly, channel 1 (also not shown) is hardly distinguishable
	from either of the other two.
\label{fig:focus_series}
}
\end{center}
\end{figure*}

The S, D, and K parameters were averaged over all stars in the same
region and plotted against focus position.  This process produces plots
of S, D, and K for every region of every channel at every other focus
position between 45 and 70.  At perfect focus, S should be at its
maximum while D and K should be at a zero crossing, meaning that stars
are small and circular.  In reality, due to optical aberrations and the
imperfect perpendicularity of the detector to the optical axis,
different regions of the image come into focus at different positions
and D/K do not cross zero at exactly the same position as S reaches
maximum.  Additionally, due to chromatic aberration of the lens,
optimal focus is different for the different channels.  These effects
can be seen in \reffig{focus_series}.

As a compromise between the different image regions and channels, we
chose ``58'' as the target focus position.  The actual focus position
can change with time for a variety of reasons, including physical
shrinking or expansion due to temperature changes or gravity pulling on
the lens as it changes orientation.  To compensate for these effects,
an autofocus script monitors the incoming focus images every night.  It
detects stars in the images, finds the S, D, and K parameters for all
regions and all channels, and determines the current focus position
with respect to the focus series.  The script then changes the focus by
the appropriate amount to bring the actual focus close to that
represented by position ``58'' on February 18.

Unfortunately, the autofocusing script had a software bug, and it did
not run correctly during the first month.  As a consequence, the focus
position never changed between February 24 and April 20.  Remarkably,
the lens is so stable with respect to thermal changes that even a month
and a half later, its full width at half maximum (FWHM) was still close
to 2 pixels---the same value it had been since February 24.  Even on
the worst nights, it did not change by more than $\sim$0.3 pixels. 
Ellipticity is more variable from frame to frame and throughout the
night, but seems to have remained around 0.1-0.2 across all nights of
observation.  This may be affected by tracking errors as well as
focusing errors.  The autofocus script was fixed on April 21.  The fix
did not seem to have a major impact on our results.

\subsection{PSF broadening}
\label{subsec:psf_broadening}
While the absolute pointing of the HAT mounts is relatively poor (tens
of arcminutes), the precision of tracking and fine motion of the HAT
mount are very good, usually better than the angular scale of one pixel
of the DSLR and the Apogee CCD on HAT10.  The mount is also capable of
deliberately introducing tiny tracking errors in order to broaden the
PSF.  In this mode, called ``PSF-broadening'' or ``Drizzle''
\citep{hatnet}, the sky around the field center is divided into a grid. 
The mount points to the center of one grid cell, lingers for a certain
amount of time, moves to the center of the next grid cell, and repeats
this process for the entire grid.  The geometry of the grid, and the
time that the mount spends at certain gridpoints is such that the stars
are broadened into a wider, approximately Gaussian stellar profile.

This process enlarges the PSF by a tiny amount (by a factor of 1.5)
close to homogeneously across the field, thus avoiding many of the
problems associated with defocusing while decreasing the effects of
subpixel structure and forcing the PSF to be more Gaussian than it
would otherwise be.  On HATNet CCDs, drizzle improves photometric
precision substantially.  As shown later, the same is true for the DSLR
experiment.

Field G180 was observed in D (Drizzle mode) and T (Tracking, no
drizzle) mode, alternating every frame.  This is to determine the
effect of drizzle on photometric precision, and whether high precision
can be achieved without it.

\subsection{Current data}
All frames were separated by night of observation on the HAT10 control
computer, and downlinked to a Princeton-based server on a regular
basis.  From February 22 to May 30 inclusive, we took 4645 object
frames of field G180, 1162 dark frames, 337 pointing frames, 2140
flatfields, 1445 focus frames, and 20 biases.

\section{Software}
The HAT10 hardware (mount, dome, cameras) is controlled by a PC running
Linux.  An observer daemon, written in TCL, is the master process that
controls observations.  It communicates with the CCD server, which is
responsible for communicating with the CCD and DSLR through drivers. 
After each exposure, other components of the control software append
metadata to the FITS header and record the exposure in a database.  An
autofocus script checks stellar PSFs on focus frames to correct focus,
while another script solves images astrometrically and uses the
solutions to correct pointing errors.

To communicate with the camera, our CCD server uses the {\tt
libgphoto2}
library\footnote{\url{http://www.gphoto.org/proj/libgphoto2/}}.  This
is an open-source C library, supporting a large number of DSLRs, that
abstracts away the camera protocol to provide a simple API for camera
control.  Because the communication protocol with the Canon 60D DSLR is
proprietary, and because Canon does not provide Linux drivers to
control their cameras, {\tt libgphoto2} drivers were developed by
reverse engineering the protocol.  To decode the raw ``CR2'' files, we
invoke the {\tt dcraw}
program\footnote{\url{https://www.cybercom.net/~dcoffin/dcraw/}} by
David Coffin, which we modified to also extract the camera sensor
temperature.

\subsection{Focusing}
HATNet lenses are focused by adjusting their focusing ring via a belt
and a stepper motor, controlled by the instrument control computer. 
For the DSLR, we exploited the focusing capability built in to the
DSLR, and with top-level control incorporated into our camera control
server.

The {\tt libgphoto2} library does not allow retrieval of the actual
focus position, nor does it allow focusing to an absolute position. 
Instead, it allows one to move the focus by a small, medium, or large
amount in both directions.  To overcome the limitations, we always move
the lens by a small amount every time.  Position 0 is defined as the
farthest possible focus (hitting the end position), and position N is
reached by moving the focusing nearer by N small steps from the
farthest possible focus.  The focus position was initialized by moving
the focus outwards a large number of steps so as to hit the physical
end position, then re-setting our internal counter to zero, and then
keeping track of all the steps we moved.  We assumed that there is no
drift and backlash, e.g.~moving -5 steps followed by +5 steps takes us
back to the same exact position.

\section{Image reduction}
The reduction pipeline mostly, but not entirely, follows the pipeline
used for HATNet and HATSouth \citep[see, e.g.][]{hat_p_11b}.

\subsection{Calibration of Images}
The first step in image reduction consists of bias subtraction, dark
subtraction, and flat fielding.

To perform bias subtraction, we ignore overscan regions, reject outlier
frames, and take a pixel-by-pixel median of the other frames.

To perform dark subtraction, we find groups of dark frames separated by
less than 5 degrees centigrade.  The largest such group is used to
generate master dark 1, the second largest group is used to generate
master dark 2, and so on, until no group exists with more than 10
frames.  Each group of frames is then bias-subtracted and combined.  As
far as we can tell, the DSLR cannot take an exposure without opening
the shutter.  It therefore did not take nightly darks.

We produce master flats for every clear night by imaging the twilight
sky.  Cloudy frames are rejected, and the rest are median combined
after smoothing (each color channel separately).

All science frames are then bias-subtracted, dark-subtracted, and
divided by the master flat frame.  We mark pixels above 13000 as
saturated.  Finally, the image is cropped, and we only keep the region
of $X \in [71,2671]$ and $Y \in [0,1733]$, which excludes the border
pixels (Fig.~\ref{fig:sensor_layout}).

\subsection{Astrometry}
A pre-requisite of precise photometry is astrometry, which consist of
finding the sources, identifying them with an external catalogue,
including deriving the transformation between the $X,Y$ pixel
coordinates and the RA,DEC celestial coordinates.  Optionally, the
coordinates from the external catalogue are projected back on the frame
to derive better centroid positions for the stars.

We did astrometry with {\tt solve-field} from {\tt astrometry.net}, an
open-source and highly robust blind solver \citep{astrometry_net}. 
Before solving any field, one needs index files, which describe
quadrangles of catalogue stars.  We used pre-generated index files for
the 2MASS catalogue \citep{two_mass}.

We invoked {\tt solve-field} twice.  On the first pass, the program
detects stars using the built-in star detector, {\tt simplexy}, which
assumes Gaussian PSFs.  It finds a WCS transformation between image
coordinates and RA/Dec, outputting several files, including an {\tt
.axy}, {\tt .corr}, and {\tt .wcs} file.  The {\tt .axy} 
file contains the measured positions and fluxes of detected stars.  The
{\tt .corr} contains cross-matches between detected and catalogue stars,
including the $x$ and $y$ positions for detected stars, RA and DEC
coordinates, and the pixel positions, as transformed back from the RA
and DEC coordinates of the external catalogue.  Finally, the {\tt .wcs}
file contains polynomial coefficients for the transformation (and its
inverse) from $X,Y$ pixel coordinates to RA,Dec celestial coordinates.

On the first run, {\tt solve-field} only tries  to find where the field
is, and does not attempt to optimize the WCS transformation.  It only
matches as many quadrangles as necessary to confirm this location,
yielding a rough transformation.  To derive a better transformation, we
run:
{\tt solve-field --continue -t 3 -q 0.01 -V filename.wcs filename.fits}

This ``verification'' stage tries to correlate all stars from all index
files with detected stars, and use the result to derive a better
third-order WCS transformation (-t 3).  We use a minimum quadrangle
size of $0.01 \arcmin$, as opposed to the default of $0.1 \arcmin$, so
that every index file is examined.

To decide which polynomial order to use for the spatial transformation
(between the pixel and celestial coordinates), we plotted the
difference between the detected star positions and the catalogue star
positions, transformed back into $X,Y$ using the derived fit.  The
standard deviation of this difference is 0.14\,pix for $X$, 0.18 for
$Y$.  There are a few extreme outliers, but when those are excluded,
the standard deviation goes down to 0.12 pixels for both axes.  We
plotted $\delta x$ and $\delta y$ against $x$, $y$, $x \times y$, and
$x^2 + y^2$, and found no dependence on any quantity for $t=3$, but a
clear dependence for $t=2$, i.e.~confirming that $t=3$ is adequate. 

We then run {\tt solve-field} on all object frames that were not
aborted mid-exposure.  Astrometry was run on each channel separately. 
Altogether, astrometry succeeded for more than 99\% of the frames.

Those few frames where astrometry failed had extreme tracking errors
due to temporary malfunctioning of the telescope mount. 

\subsection{Catalog projection}
To decide which stars to do photometry on, we query the 2MASS catalogue
down to an (extrapolated) V band magnitude of 14.

The celestial coordinates of the catalogue stars are transformed to pixel
coordinates for every science frame using {\tt wcs-rd2xy}, a tool
provided by {\tt astrometry.net}, and the {\tt .wcs} file corresponding
to the frame.

\subsection{Aperture photometry}
We used the standard aperture photometry routines of HATNet.  The
precision of aperture photometry depends heavily on the aperture radius
used to measure total flux.  For HATNet CCDs, for example, the RMS
values of light curves exhibit oscillatory behavior as a function of
radius at small radii, and start decreasing slowly but monotonically at
large radii.  On an RMS vs.~aperture graph, the peaks and troughs at
small radii differ by as much as a factor of two.  This is probably due
to a combination of not accounting properly for the non-uniform
distribution of light across each pixel and subpixel sensitivity
variations.

We use 3 apertures to perform photometry, because the optimal aperture
depends on the brightness of the star; large apertures for bright
stars, and small apertures for faint stars, where the sky background is
high compared to the flux of the star.  After testing 10 apertures from
1.2 to 3 pixels, we chose 1.4, 1.6, and 1.8 pixels as the final
apertures. 

\subsection{Magnitude fitting}
The raw magnitudes from aperture photometry for a selected star
typically exhibit large variations, sometimes more than a magnitude as
a function of time.  This can be due to many factors, including
atmospheric extinction and changing PSF shape.

Many other factors affect the raw magnitudes.  For example, as a star
drifts across the image, its center lands on different positions of the
Bayer filter and changes the amount of light falling on the sensitive
portion---an effect which our aperture photometry does not correct for. 
We employed ensemble magnitude fitting to correct for these effects.

To perform magnitude fitting, we chose a ``reference'' frame for each
tracking mode (simple tracking and PSF broadening) and each channel. 
The ``reference'' frame is chosen to be extraordinarily good: low sky
background, high S/N, field close to culmination, etc.

Every other frame is compared to the
reference frame, and the difference in raw magnitudes is computed for
every star.  These differences, $\Delta m \equiv m_i - m_{i,ref}$ for
each $i$ star, are fit by a polynomial
that depends on position, catalog magnitude, catalog J-K color, and
subpixel position.  The contribution of the position to the polynomial
is itself a fourth-order polynomial.  

The contribution of catalog
magnitude is of the form:
\begin{align*}
\sum_{i=0}^2\sum_{j=0}^2\sum_{k=0}^2c_{ijk}M^ix^jy^k
\end{align*}
In other words, it is the product of a second-order polynomial in
position and a second-order polynomial in magnitude.  This is purely
empirically motivated.  Similarly, the J-K and subpixel position
contributions, also empirically motivated, are both the product of a
first-order polynomial in J-K (or subpixel position) and a second-order
polynomial in position.  Since the four channels and two tracking modes
produce different images, we run magnitude fitting separately on all
eight combinations of channel and tracking mode.  The ensemble
magnitude fitting is performed in an iterative way, by replacing the
weights for each star in the fit with the derived r.m.s.~of the light
curves, and the magnitude of the stars with mean magnitude of the light
curves.

\subsection{Light curves}
Light curves were derived from the magnitude-fitted photometry files. 
Light curves with fewer than 1000 measurements were discarded.  These
are usually stars very close to the border of the frame that
accidentally wandered into the frame due to pointing jitter,
esp.~present at the first few exposures and initial pointing of the
telescope. 

Our light curves contain measurements for all 3 apertures used in
aperture photometry.  As expected, bright stars do better with larger
apertures and dim stars do better with smaller apertures, with the
lowest-deviation light curves having an optimal aperture of 1.8 pixels.

\subsection{EPD, TFA}
The final stages of noise reduction are External Parameter
Decorrelation \citep[EPD;][]{hat_p_11b} 
and the Trend-Filtering Algorithm \citep[TFA;][]{tfa_algorithm}.  We use
{\tt vartools} \citep{vartools} to perform both tasks.  

EPD involves finding correlations between the magnitude and so-called
external parameters.
These correlations are then modeled using a simple function, and once
the parameters are fitted, the function is subtracted from the light
curve.  External parameters include subpixel position, zenith angle,
PSF parameters, and hour angle.

We examined the dependence of magnitude on all 7 external parameters
mentioned above by plotting magnitude against the parameter for sample
light curves, as well as by fitting functions between these parameters
and stellar magnitudes to see how much the RMS error decreases.  Out of
all external parameters, subpixel position had the strongest influence. 
Even though subpixel position was a parameter in ensemble magnitude
fitting, the parameter was only fit to first order.  We found that a
sinusoidal function with a period of 1 pixel is a better fit for this
relationship than any polynomial of $x$ and $y$ up until fourth order. 
However, after subtracting the sinusoidal fit, there still remains a
slight linear relationship between magnitude and subpixel y position. 
We therefore also subtract off a linear term proportional to the
subpixel y position.  We stress that this is done for each star, 
i.e.~each light curve, separately.

Out of the remaining parameters, only the zenith angle (Z), PSF S
parameter, and hour angle (h) had any non-negligible correlation with
magnitude.  We simply fit them with linear functions.  The improvement
in RMS error after introducing a quadratic term in addition to the
linear for any of these parameters was negligible.

The final function we fit for is:
\begin{align*}
mag = a_xsin(2\pi x)+b_xcos(2\pi x)+a_ysin(2\pi y) + \\
b_ycos(2\pi y) + a_{y2}(y-floor(y)) + a_zZ + a_sS + \\
a_hh + const
\end{align*}
For every light curve we first reject outliers that are more than 5
standard deviations away from the mean.  This process is repeated 3
times, after which we fit the function above to the magnitude, subtract
off that function for every data point, and apply a constant
zero-point, to match the standard magnitude of the star in the
catalogue.  Besides de-correlating for external parameters, this also
transforms raw instrumental magnitudes to the standard system.

\begin{figure*}[t]
\includegraphics[width=\textwidth]{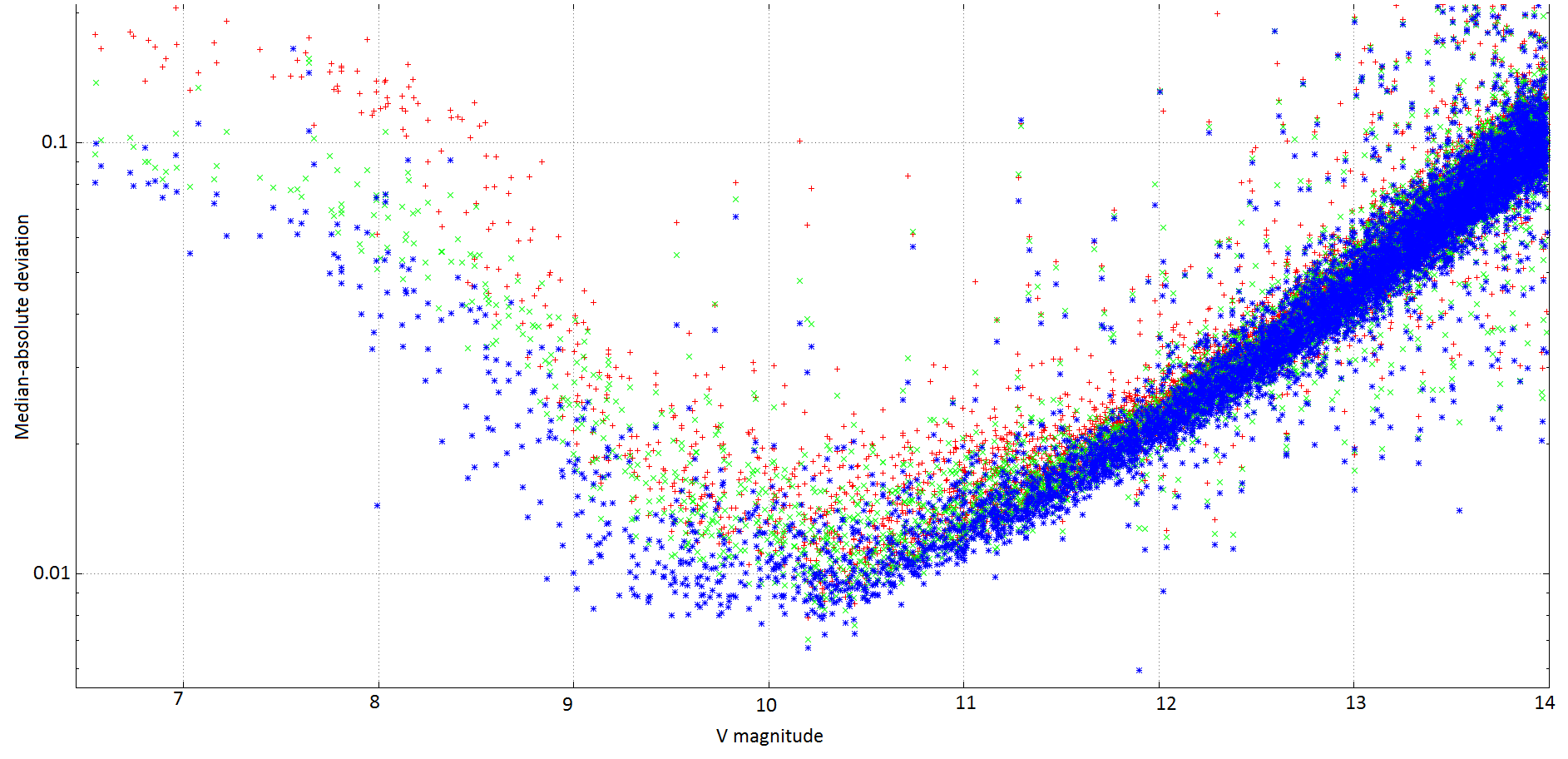}
\caption{
	Median absolute deviation vs.~magnitude for channel 0
	(corresponding to green light) before EPD
	(indicated by green crosses), after EPD but before TFA (red plus
	symbols), and after TFA (blue star symbols). 
	This plot is for D20 tracking mode.
}
\label{fig:epd_tfa_effect}
\end{figure*}

After EPD, we perform the trend-filtering algorithm
\citep{tfa_algorithm} to remove remaining systematic variations.  We
use the {\tt vartools} implementation for this task.  For every light
curve, {\tt vartools} retrieves the magnitudes for the template stars
and fits a linear function, modeling the light curve as a linear
combination of the template light curves.  Template stars less than 25
pixels away from the target star are excluded from the fit.  This
prevents an artificially good fit between stars whose PSFs are blended
together, and also prevents template stars from being fit to their own
light-curves.

The template stars were selected by hand.  First, the image region was
divided horizontally and vertically into 3 equal portions, for a total
of 9 regions.  From each image region, 11 stars were selected, roughly
2 for each magnitude bin between $r = 8\ldots13$.  We picked the stars
with the smallest median absolute deviations (MADs), and did not select
any star that had a MAD of more than 5 percent for channel 0 in D20
mode.

For HATNet Apogee CCDs, TFA drastically reduces the RMS error, from a
best-case accuracy of 8 mmag to 5 mmag.  This is also true for the DSLR
observations.  \reffig{epd_tfa_effect} shows the effect of EPD
and TFA on DSLR images.  Notice that TFA makes a large difference to
the precision of results, whereas EPD makes a small difference.

\subsection{Transiting planet candidate search}
All three months of data were searched for transit candidates, and four
candidates were found in field G180.  One of them is KELT-3b, the only
known transiting exoplanet in the field, with a transit depth of
0.89\%.  Unfortunately, follow-up observations show that all three of
the other candidates are false alarms.  In two, the photometric signal
(which was weak to begin with) proved to be an artifact; in the other,
the star was a periodic variable.

\section{Results}

At the end of the reduction pipeline, light curves were derived for
every $V < 14$ catalogue star within the frame.  To demonstrate the
photometric precision we achieved, we plotted the MAD for every star
against the appropriate catalogue magnitude for that star.  MAD is the
median of the absolute value of the difference from the median.  This
statistic is used because it is mostly insensitive to outliers, and we
did not want the results to depend on which algorithm was used to
detect and reject outliers.  Note that MAD is smaller than the
standard deviation by a factor of 1.48 for Gaussian distributions.

\begin{figure*}[t]
\subfigure[]{
	\includegraphics[width=0.5\textwidth]{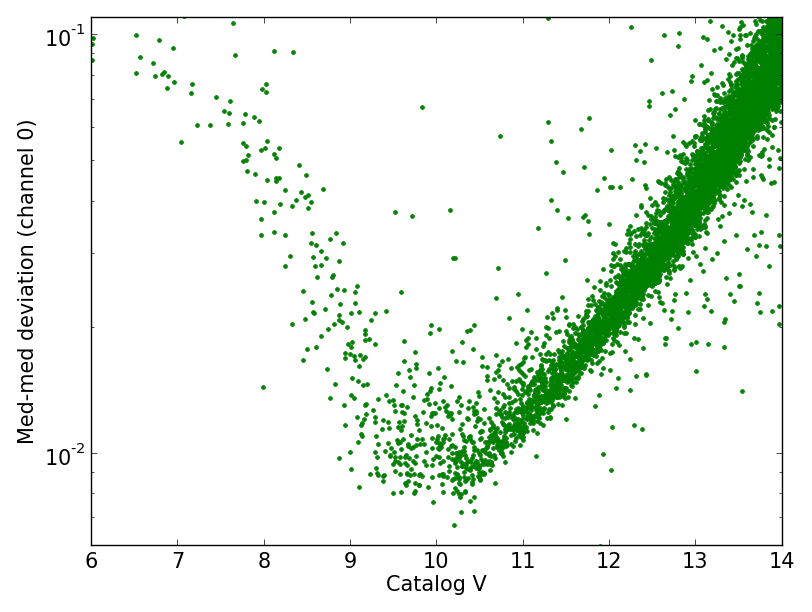}}\qquad
\subfigure[]{
	\includegraphics[width=0.5\textwidth]{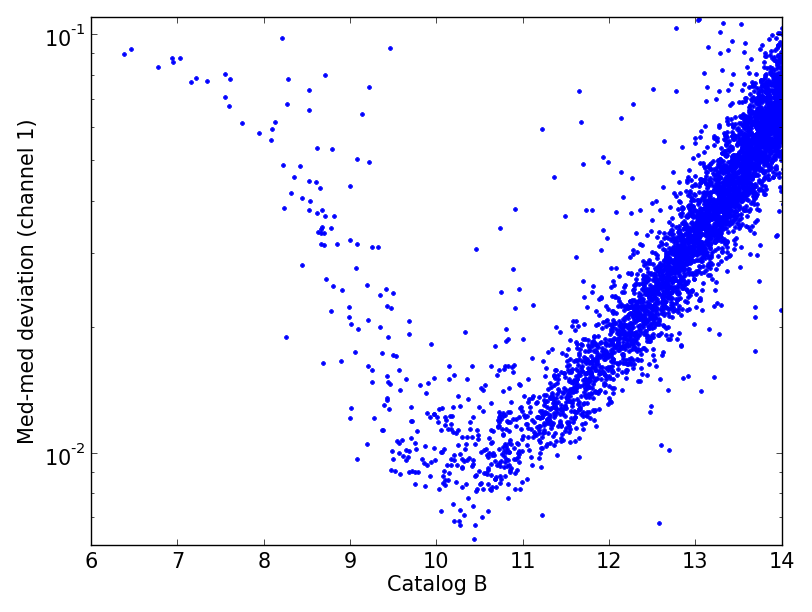}}\\
\subfigure[]{
	\includegraphics[width=0.5\textwidth]{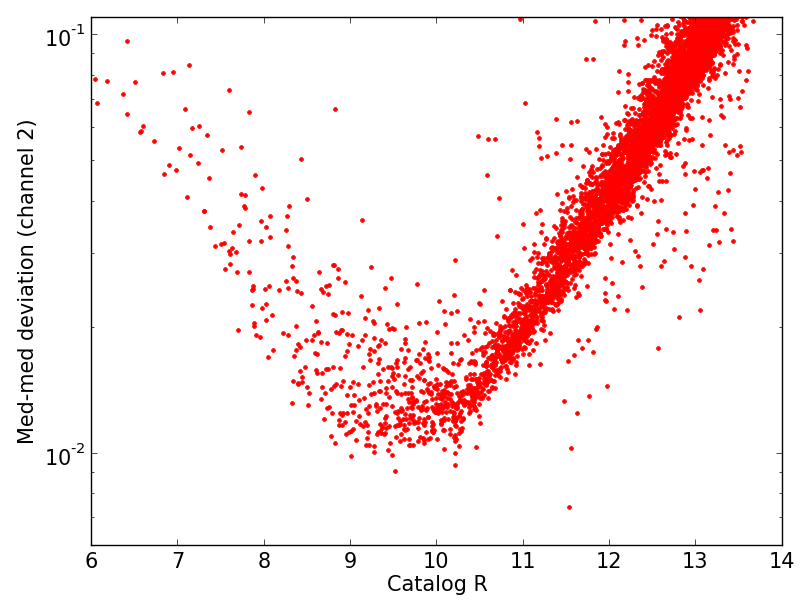}}\qquad
\subfigure[]{
	\includegraphics[width=0.5\textwidth]{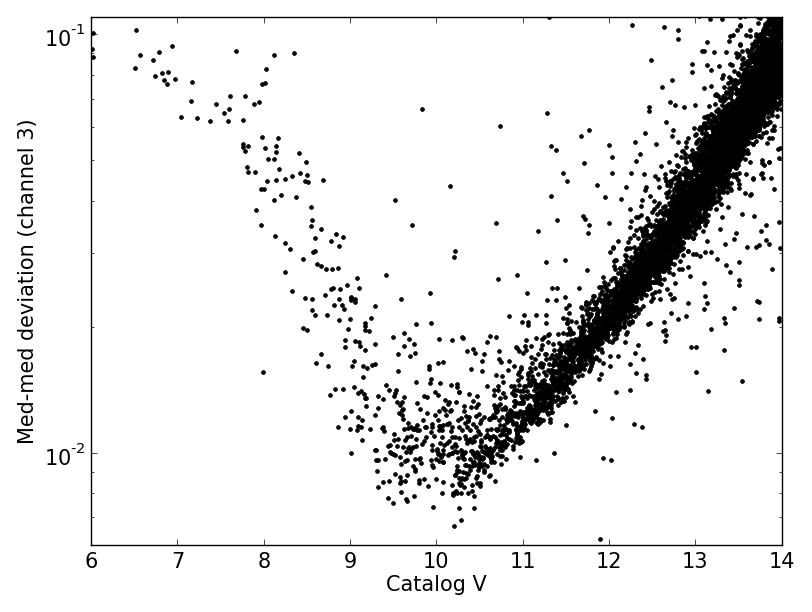}}
\caption{
	Median absolute deviation for all four channels, with D20 tracking
	mode.
}
\label{fig:D_precision}
\end{figure*}

\begin{figure*}[t]
\subfigure[]{
	\includegraphics[width=0.5\textwidth]{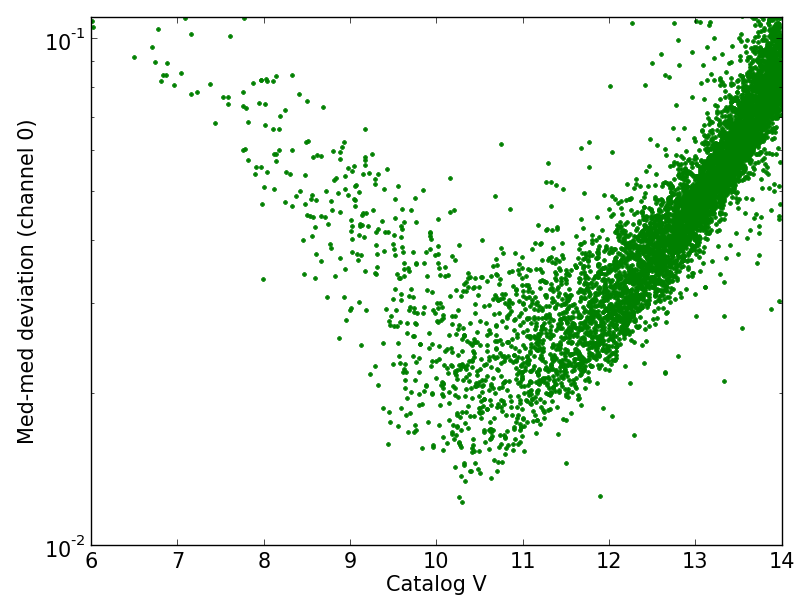}}\qquad
\subfigure[]{
	\includegraphics[width=0.5\textwidth]{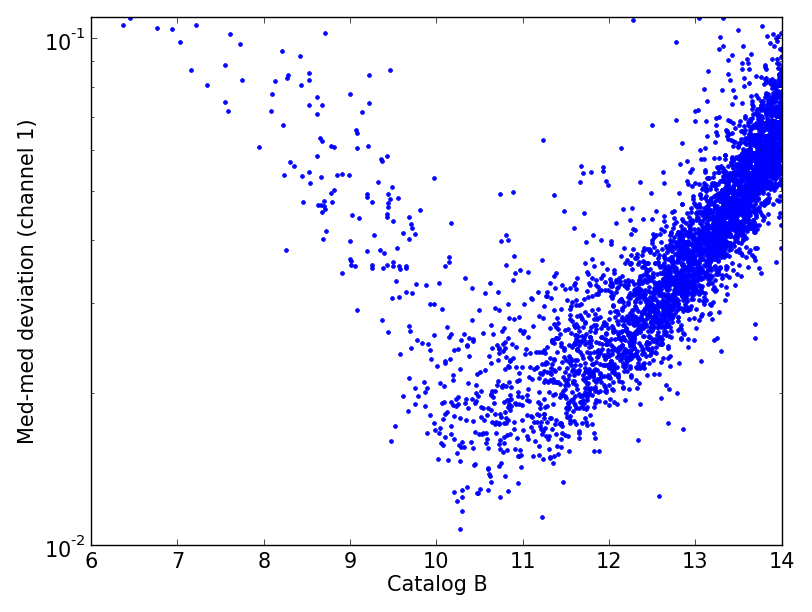}}\\
\subfigure[]{
	\includegraphics[width=0.5\textwidth]{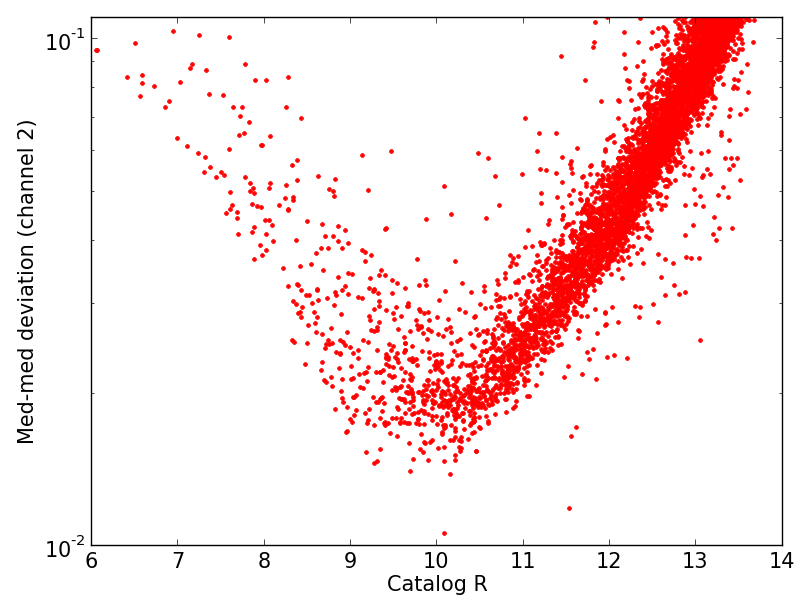}}\qquad
\subfigure[]{
	\includegraphics[width=0.5\textwidth]{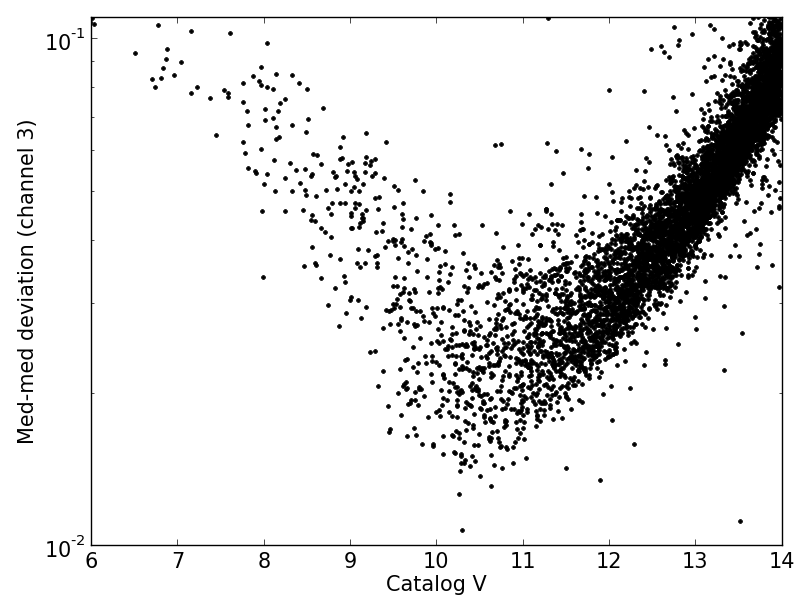}}
\caption{
	Median absolute deviation for all four channels, with T tracking mode.
}
\label{fig:T_precision}
\end{figure*}

\begin{figure*}[h]
\subfigure[]
	{\includegraphics[width=0.45\textwidth]{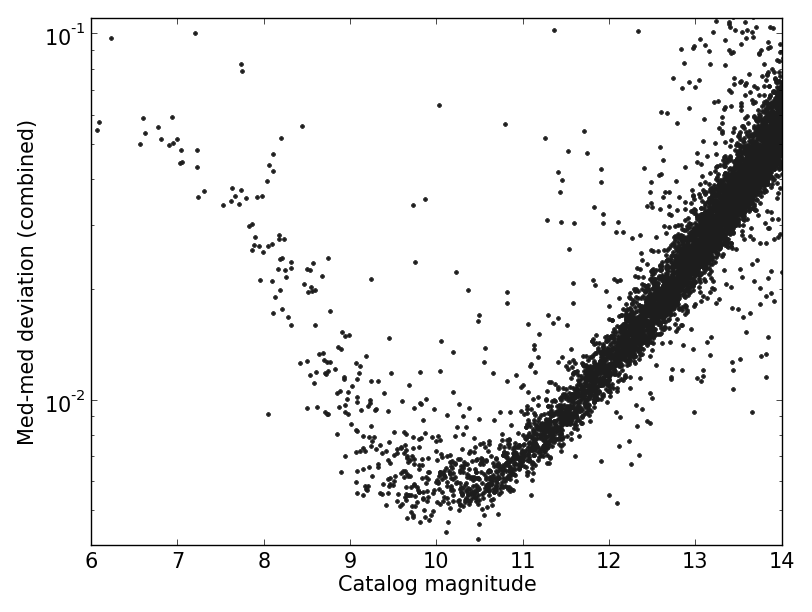}}
\subfigure[]
	{\includegraphics[width=0.45\textwidth]{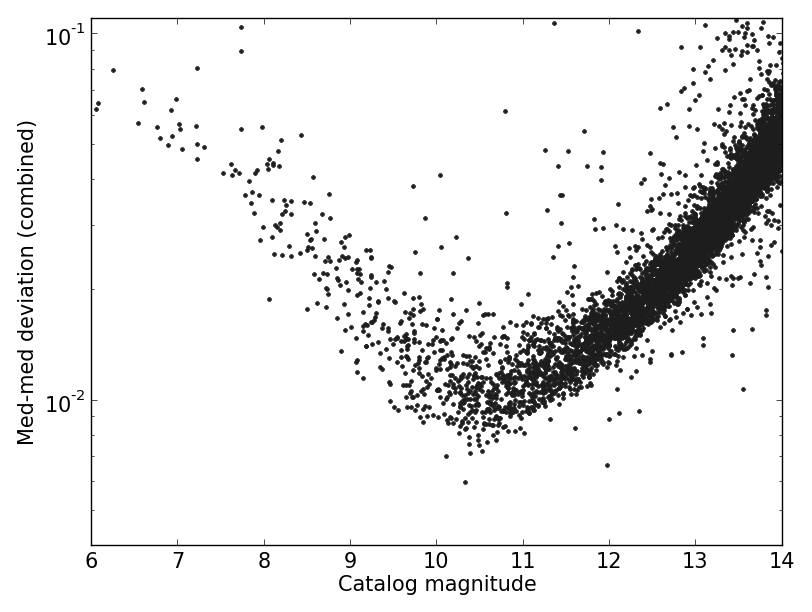}}
\caption{
	Median absolute deviation for all four channels combined.  Left: D20
	(drizzle); right: T (tracking).
}
\label{fig:combined_precision}
\end{figure*}

The results are shown in \reffig{D_precision} for D20 tracking
mode and \reffig{T_precision} for T tracking mode.  
\reffig{combined_precision} shows the precision of all four channels
combined.  The combining was done by weighting every channel by the
inverse square of the channel's MAD, and adding the results.  We have
also tried combining the channels in the simplest possible way: by
adding the four individual channel magnitudes and dividing by four,
which is equivalent to taking the geometric mean of the fluxes. 
Surprisingly, the results from simple addition and from weighted
addition are almost exactly the same: in a deviation vs.~magnitude
graph there is almost no discernible difference between the two.

The shape of these graphs is very typical of photometric measurements. 
Faint stars have a high MAD because sky noise, photon noise, read
noise, and dark noise all dominate over the signal.  Very bright stars
also have a high MAD because they saturate the sensor, making it
impossible to precisely measure their flux.  In between the two
extremes is a minimum where photometric precision is the best.

As can be seen, D20 (PSF broadening) is far better than T (tracking)
for all channels.  In addition to having a lower minimum, the envelope
is also narrower, resulting in more stars measured at a better
precision.

The combined magnitudes for D20 tracking mode have a minimum med-med
deviation of 4.6 mmag, and 1000 stars between magnitudes 9 and 11.5 are
measured more accurately than 10 mmag.  For T tracking mode, the
combined magnitudes have a minimum deviation of 8 mmag, with far fewer
stars lower than 10 mmag.  For the individual channels, 0, 1, and 3
have a minimum deviation of 7-8 mmag for D20, and roughly 15 mmag for T
mode.  Channel 2 is substantially worse than the others, with a minimum
deviation of 10 mmag for D20 and 17 mmag for T.  This is not
surprising, considering that the red channel has a lower quantum
efficiency than the others.

\subsection{Sample light curves}
To further demonstrate the precision of our photometry, we analyzed the
light curves for selected variable stars. 

\begin{figure*}[t]
\subfigure[]
	{\includegraphics[width=0.5\textwidth]{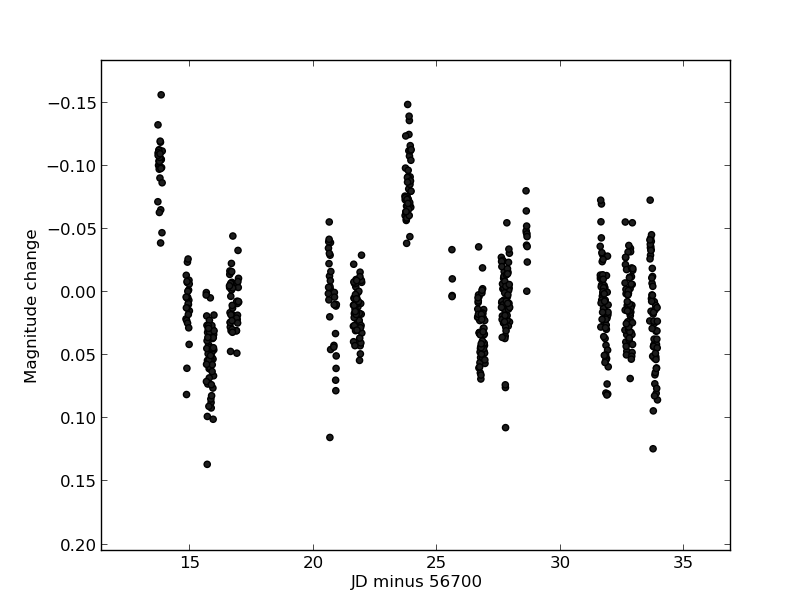}}
\subfigure[]
	{\includegraphics[width=0.5\textwidth]{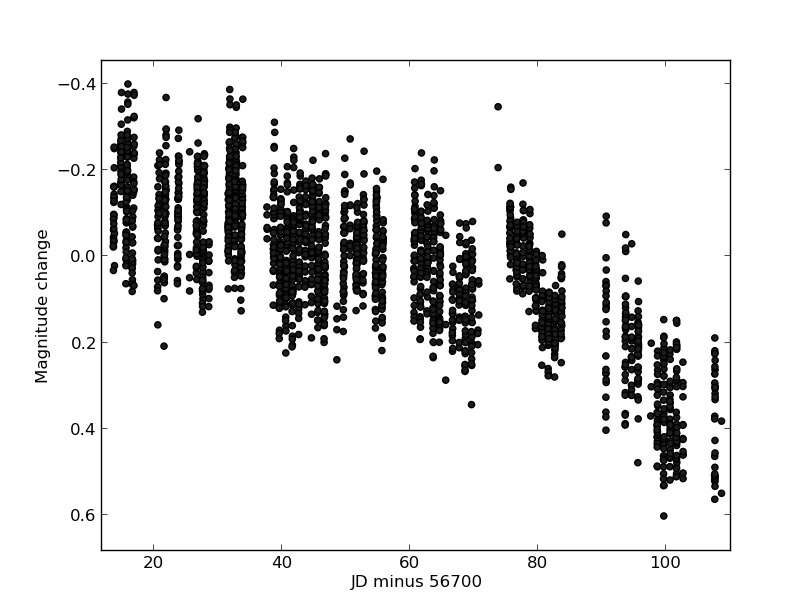}}
\caption{
	Light curves for a RS Canum Venaticorum-type variable (HAT-180-0000594,
	left) and a Mira-like late-type giant (HAT-181-0000002, right).
	Color channels have been combined for both.
}
\label{fig:irregular_variables}
\end{figure*}

In \reffig{irregular_variables}, we plot the magnitude difference from
median magnitude against time for two variable stars.  On the left is
ZZ LMi, an RS Canum Venaticorum-type variable, characterized by strong
chromospheric activity.  Like solar activity, there is no strong
periodicity to the light output of this variable.  According to the
General Catalog of Variable Stars, this star varies between magnitude
10.8 and 12.1.  However, the amplitude of variations may change over
the years, due to cycles similar to the solar cycles.

In the right panel of \reffig{irregular_variables} is U LMi, a
semiregular late-type giant, similar to Mira but with smaller
variations.  This star varies between magnitude 10 and 13.3 with a
272.2-day period, but we only see a secular upwards trend over our
observations.

\begin{figure}[H]
\includegraphics[width=0.5\textwidth]{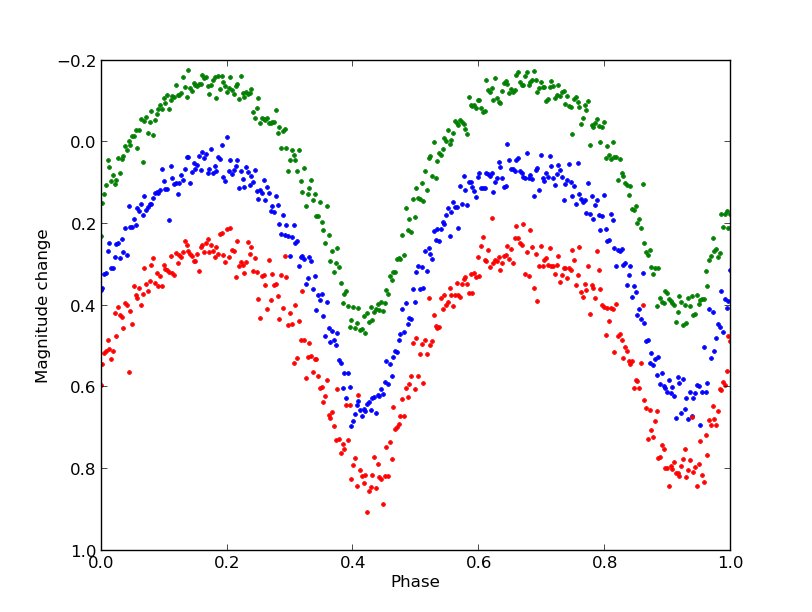}
\caption{
	Light curves for an eclipsing contact binary, HAT-180-0001149,
	phase-folded to the orbital period and binned into bins of one
	degree.  The colors are arbitrarily offset by 0.2 mag.  
}
\label{fig:RT_LMi}
\end{figure}

\reffig{RT_LMi} shows the light curve for RT LMi, 
an eclipsing contact binary with a period of 0.3749 days
and a range of magnitude variation of 11.4--11.7.  

\begin{figure}[h]
\includegraphics [width= 0.5\textwidth]{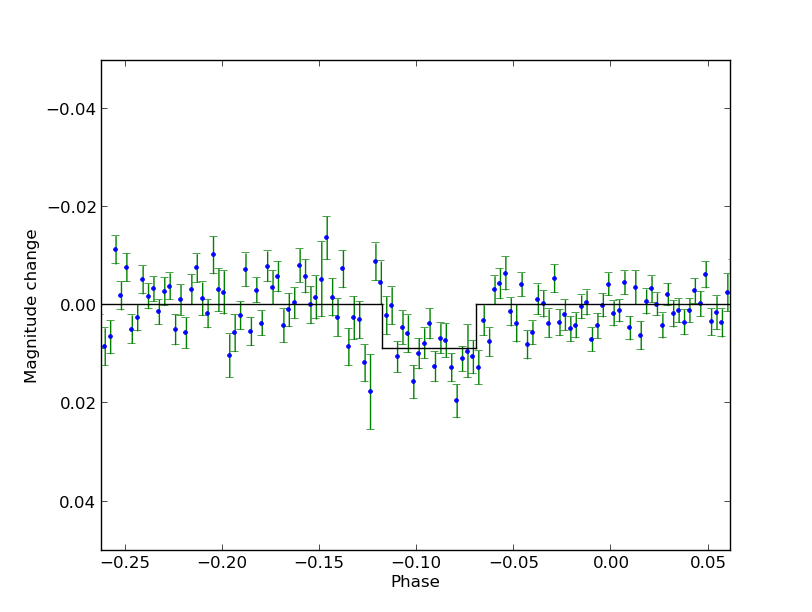}
\caption{
	Phase-folded and phase-binned light curve for KELT-3. Color
	channels have been combined.
}
\label{fig:kelt3b}
\end{figure}

Our automatic search pipeline detected KELT-3b, a known transiting
extrasolar planet \citep{kelt3b}.  KELT-3b has a period of 2.7 days, a
transit duration of 189 minutes, and a transit depth of 8.8 mmag.  The
light curve is shown in \reffig{kelt3b}, where it is phase-folded and
phase-binned into one-degree bins.  The transit signal can clearly be
seen.  Note that this result is still suboptimal because KELT-3b is
only 20 pixels away from the image edge, leading to a distorted PSF
shape.

\section{Comparison with the CCD}

\begin{figure*}[t]
\subfigure[]
	{\includegraphics[width=0.5\textwidth]{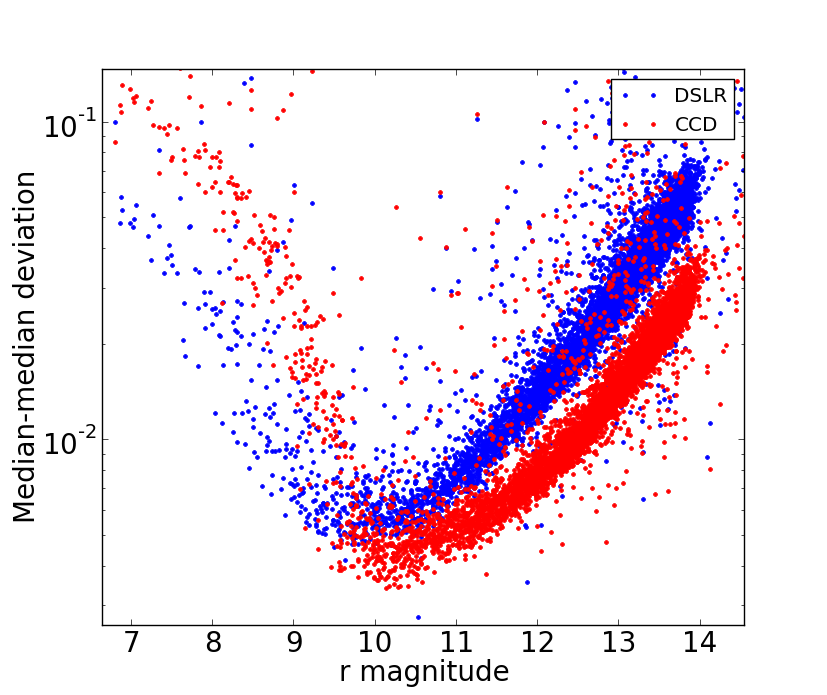}}
\subfigure[]
	{\includegraphics[width=0.5\textwidth]{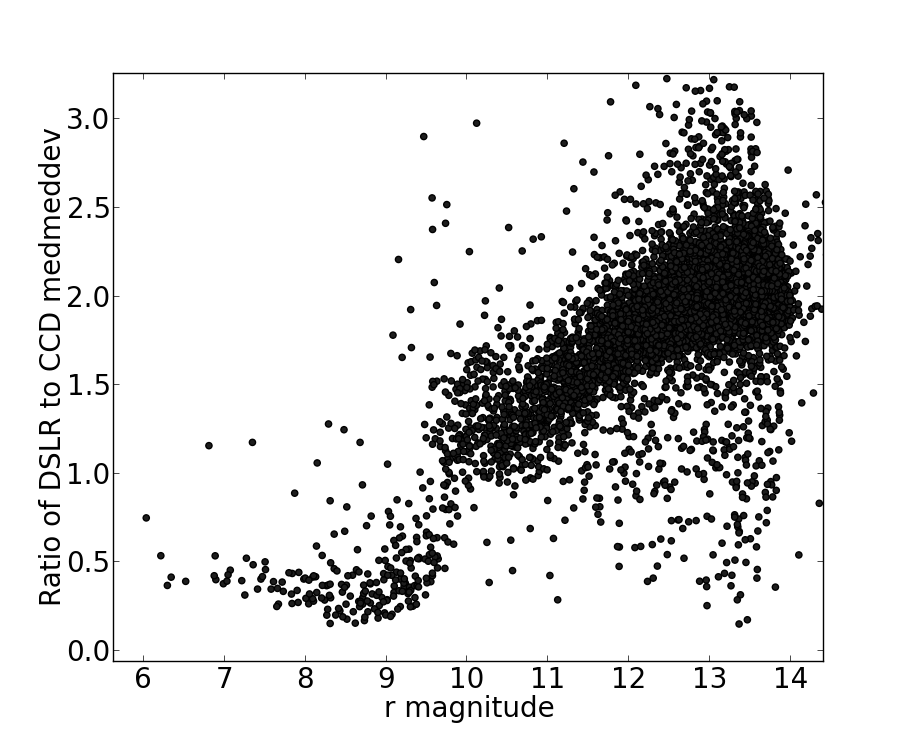}}
\caption{
	Left: Median absolute deviation of the CCD and DSLR for all stars
	in G180, plotted against the $r$ magnitude of the stars.  Right:
	For every star, the ratio of its DSLR median absolute deviation to
	its CCD MAD is calculated and plotted as a function of the stellar
	$r$ magnitude.  The jump at $r\approx10$ is due to the CCD
	saturating for bright stars, where the DSLR still produces
	reasonable photometry.
}
\label{fig:ccd_dslr_comp}
\end{figure*}

To compare the DSLR with the CCD, we did a parallel reduction of the
HAT10 CCD data for exactly the same time instances, i.e.~under the same
exact observing conditions.  \reffig{ccd_dslr_comp} compares the MAD of
both the DSLR and the CCD, and in the same figure, we also show the
ratio of the deviations.  The CCD achieves a minimum deviation of 3.4
mmag, with 3377 stars measured more accurately than 10 mmag.  As
mentioned before, these numbers are 4.6 mmag and 1000 stars for the
DSLR, respectively.  The spread of the magnitude-deviation relationship
is very similar for both devices.  Notice also that the CCD becomes
imprecise very quickly at magnitudes lower than the saturation limit,
whereas the DSLR becomes imprecise more gradually.  This partially
explains the knee in \reffig{ccd_dslr_comp}, the other part of the
explanation being the difference in saturation magnitude between the
devices.

Time-correlated (``red'') noise is also very important in the detection
of signals.  Notably, red noise decreases the transiting planet
detection efficiency, and often leads to catching false signals.  To
estimate the variance due to time-correlated noise, we measured the
autocorrelation of the light curves and modeled it using a function of
the form:

\begin{align*}
r = Ae^{-t/\tau}
\end{align*}
where t is the lag and A is the fraction of the variance caused by red
noise, so that $RMS_\mathrm{red}/RMS = \sqrt{A}$.  To compute A, we
derived the auto-correlation function for all light curves in Drizzle
mode, for lags between 0 and 0.1 days at intervals of 5 minutes.  The
sequence of autocorrelations for each light curve is normalized to 1
for lag 0.  Autocorrelations are then divided into bins according to
the V magnitude of the star, and merged.  The bins are one magnitude
wide and cover the magnitude range of 8 to 13.

\begin{figure}[h]
\includegraphics [width= 0.5\textwidth]{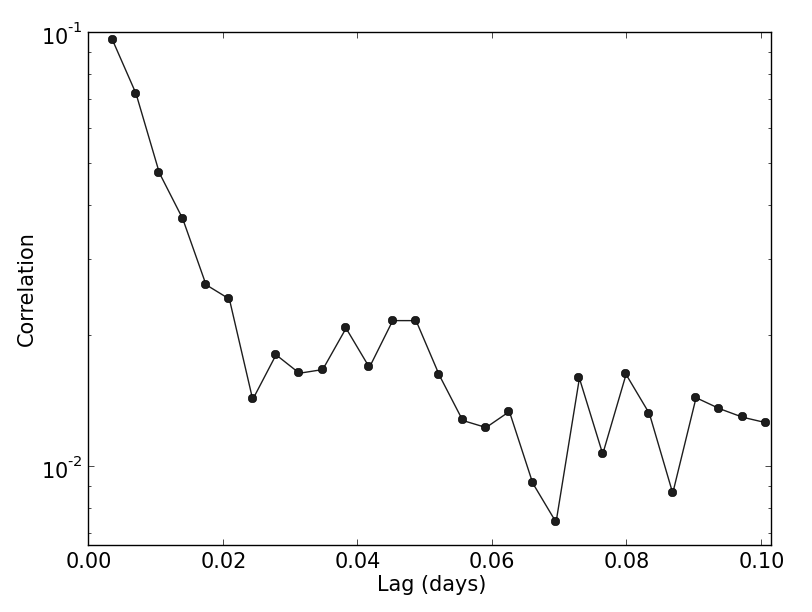}
\caption{
Combined autocorrelation for stars of r-band magnitude 10-11.
}
\label{fig:autocorr_dslr_10_11}
\end{figure}

\reffig{autocorr_dslr_10_11} shows the merged autocorrelations for
magnitude 10-11, but the other autocorrelations look similar.  At small
lags, the correlation decreases rapidly in a relation very close to
exponential.  At around 0.025 days, the relation changes to one that
decreases much more slowly, and is much farther from a perfect
exponential.  To calculate the red noise ($A$) we take the first 6
points (the ones with lag $<$ 0.025\,d) from the combined
autocorrelation, and fit them to this exponential form. 
Table~\ref{table:red_noise} shows the results, compared with the
results of a similar analysis on the CCD.  As expected, red noise is a
greater proportion of total noise for brighter stars.

\begin{table}
\begin{center}
\begin{tabular}{llllr}
\hline
\text{Magnitude} & $A_{DSLR}$ & $\tau_{DSLR}$(d) 
& $A_{CCD}$ & $\tau_{CCD}$(d)\\ 
\hline
8-9 & 0.38 & 0.089 & 0.60 & 0.24\\
9-10 & 0.17 & 0.033 & 0.12 & 0.38\\
10-11 & 0.12 & 0.028 & 0.068 & 0.32\\
11-12 & 0.11 & 0.024 & 0.057 & 0.33\\
12-13 & 0.08 & 0.022 & 0.045 & 0.42\\
\hline
\end{tabular}
\end{center}
\caption{
	Red noise characteristics of the DSLR and CCD light curves.
}
\label{table:red_noise}
\end{table}

\section{Conclusion}

In this paper, we demonstrated the capability of DSLRs for precision
photometry.  With D20 tracking mode and a 1.4-1.8 pixel aperture, we
achieved a best-case median absolute deviation of 7-8 mmag for channels
0, 1, and 3 (representing green, blue, and green respectively) and 10
mmag for channel 2.  When these channels are combined by weighted or
simple addition, the deviation is around 4.6 mmag.  Unfortunately, it
is also true that this level of precision depends crucially on the
PSF-broadening observing technique, whereby the stellar PSFs are
broadened by a small fraction of the original PSF width.  In tracking
mode, it is only possible to achieve 10 mmag photometry or better for a
tiny fraction of stars, even at the optimal magnitude and aperture.  To
our knowledge, this is the first time that a precision significantly
better than 10 mmag has been achieved with a DSLR without extreme
defocusing, over an extended duration, and over a large field-of-view,
for a large number of stars.

This level of precision allows for DSLRs to be used in exoplanet
detection, and certainly in studying variable stars.  Our result of 4.6
mmag is not drastically worse than the 3-4 mmag of the Apogee CCDs that
HATNet currently uses.  This loss of precision is partly compensated by
the low cost of the DSLR (it is over an order of magnitude cheaper than
the CCDs), and its capability for simultaneous multiband photometry. 
Additionally, DSLRs, like most consumer electronics, are improving at a
rapid pace.  A newer DSLR may perform much better than the 60D, with
the downside of being less tested and less well-supported by existing
software.

\acknowledgements 

\paragraph{Acknowledgements}
HATNet operations have been funded by NASA grants NNG04GN74G and
NNX13AJ15G.  Follow-up of HATNet targets has been partially supported
through NSF grant AST-1108686.  K.P.~acknowledges support from NASA
grant NNX13AQ62G.  Data presented in this paper are based on
observations obtained at the HAT station at the Fred Lawrence Whipple
Observatory of SAO. We wish to thank M.~Calkins, P.~Berlind and
G.~Esquerdo at FLWO for their dedicated help in operating the HAT
telescopes. We also thank Princeton University for coordinated turning off
of the stadion lights next to Peyton hall to facilitate initial
observations. 


\bibliographystyle{apj}
\bibliography{dslr_bib}

\end{document}